\RequirePackage{fix-cm}
\documentclass[twocolumn,epjc3]{svjour3}
\smartqed  % flush right qed marks, e.g. at end of proof
\usepackage{amsmath}
\usepackage{amssymb}

\RequirePackage{graphicx}
\RequirePackage[colorlinks,citecolor=blue,urlcolor=blue,linkcolor=blue]{hyperref}
\RequirePackage[numbers,sort&compress]{natbib}
\RequirePackage{microtype}
\providecommand{\doi}[1]{DOI #1}
\renewcommand{\doi}[1]{\href{https://doi.org/#1}{DOI #1}}

\RequirePackage{mathptmx}      % use Times fonts if available on your TeX system
\journalname{Eur. Phys. J. C}
\begin{document}

\title{Scalar quasinormal and quasibound states of weakly magnetized Ernst--Schwarzschild spacetime: a confluent-Heun Wronskian approach%\thanksref{t1}
}


\author{Danilo S. Ferreira\thanksref{addr1} \and Luis B. Castro\thanksref{e1,addr1,addr2} %etc.
}

\thankstext{e1}{e-mail: lrb.castro@ufma.br}


\institute{Programa de P\'{o}s-graduação em F\'{i}sica, Universidade Federal do Maranh\~{a}o, Campus Universit\'{a}rio do Bacanga, 65080-805, S\~{a}o Lu\'{\i}s, Maranh\~{a}o, Brazil.
\label{addr1} \and
Coordenaç\~{a}o do Curso de F\'{i}sica - Bacharelado, Universidade Federal do Maranh\~{a}o, Campus Universit\'{a}rio do Bacanga, 65080-805, S\~{a}o Lu\'{\i}s, Maranh\~{a}o, Brazil.\label{addr2}
}

\date{Received: date / Accepted: date}

\maketitle

\begin{abstract}
We study the quasinormal-mode and quasibound-state spectra of a massless scalar field in weakly magnetized Ernst--Schwarzschild spacetime. In the separated $O(B_0^2)$ regime, the Klein--Gordon equation reduces to a radial confluent-Heun equation. We construct an endpoint-selected spectral Wronskian directly from the canonical confluent-Heun recurrence and its regular-to-irregular connection limit. Quasinormal modes and quasibound states are then obtained as zeros of the same analytic spectral function on the radiative and bound sheets, respectively, providing a unified treatment of the two spectral sectors within the separated weak-field problem. The resulting QNM and QBS frequencies agree with independent weak-field Ernst and massive-Schwarzschild benchmarks available in the literature. We further show, for representative cases, that the frequencies selected by the local Heun-series termination relation examined here do not satisfy the endpoint-selected quasibound-state condition. The construction therefore provides a direct horizon-to-infinity spectral condition for both radiative and bound states within the weak-field Ernst problem.

%\PACS{04.62.+v \and 04.20.Jb \and 03.65.Pm \and 03.65.Ge}
\end{abstract}

\section{Introduction}
\label{sec:introduction}

Black-hole perturbations are naturally formulated as spectral boundary-value problems. Quasinormal modes (QNMs) describe the dissipative response selected by purely ingoing behavior at the event horizon and outgoing behavior in the asymptotic region, and they play a central role in the dynamics and stability of black-hole spacetimes \cite{CQG26:163001:2009}. When an effective mass or another confining mechanism is present, the same exterior geometry may also support quasibound states (QBSs), for which the horizon condition remains ingoing while the field decays at large radius \cite{PRD76:084001:2007,PRD89:083006:2014}. Although the two spectra have different physical interpretations, both require information from widely separated regions of the radial problem. Their determination is therefore intrinsically global: a local solution selected at the horizon must be connected to the appropriate physical channel at infinity.

A particularly useful setting in which both sectors arise is the magnetized Schwarzschild solution of Ernst, which describes a Schwarzschild black hole embedded in an external magnetic field \cite{JMP17:54:1976}. The exact geometry is not asymptotically flat, and the scalar perturbation problem is not generically reducible to a single decoupled spherical-harmonic radial equation. In the weak-magnetization approximation through $O(B_0^2)$, however, the massless Klein--Gordon equation separates and the radial equation acquires the effective term $B_0^2m^2$ \cite{PLB659:375:2008,EPJC75:391:2015}. This separated model has been used to study scalar QNMs by several methods \cite{PLB659:375:2008,EPJC75:391:2015}, while Senjaya \cite{EPJC84:57:2024} subsequently analyzed quasibound frequencies through a confluent-Heun representation of the same weak-field radial equation. More recently, a charged-scalar analysis performed directly in the exact magnetized background has emphasized the coupled multipolar structure that appears beyond the separated weak-field description \cite{PRD113:124029:2026}. The $O(B_0^2)$ model should therefore be viewed as a controlled spectral problem in its own right rather than as the unrestricted exact-Ernst problem.

For fixed azimuthal number $m$, the weak-field radial equation is formally identical to that of a massive scalar field in Schwarzschild spacetime after the identification $\mu_{\rm eff}=B_0|m|$ \cite{PLB659:375:2008,EPJC75:391:2015}. This correspondence makes both radiative and bound sectors natural within the same separated equation and provides independent massive-Schwarzschild results against which a QBS construction can be tested. At the same time, existing analyses organize the two sectors in rather different ways. QNMs have primarily been obtained with Frobenius or continued-fraction techniques, WKB approximations, and time-domain calculations \cite{PLB659:375:2008,EPJC75:391:2015}, whereas the quasibound analysis of Ref.~\cite{EPJC84:57:2024} relies on a local Heun-series termination relation. This raises a more basic spectral question: can the physical QNM and QBS conditions be formulated directly from the same horizon-to-infinity connection problem, without treating them as unrelated quantization prescriptions?

The confluent-Heun structure of the radial equation provides a natural setting for addressing this question. Heun equations arise in many separated black-hole wave problems, and they have been used both to write spectral conditions directly in terms of confluent-Heun functions \cite{PRD84:127502:2011} and to construct explicit connection coefficients with applications to scattering and quasinormal spectra \cite{PRD105:044047:2022}. Most closely related to the global viewpoint relevant here, Chen et al.~\cite{PRD112:103036:2025} used analytic continuation together with confluent-Heun connection formulas to construct complete QNM spectra of type-D black holes. These developments underscore an important distinction: reducing a radial equation to Heun form identifies its local analytic structure, whereas a physical spectrum is selected only after solutions associated with different asymptotic regions are connected.

In this work we formulate that connection directly for the massless scalar field in the separated weak-field Ernst--Schwarzschild problem. After reducing the radial equation to canonical confluent-Heun form and fixing the ingoing horizon solution, we define an endpoint-selected spectral Wronskian between that solution and the physical asymptotic channel. The required regular-to-irregular connection coefficient is then evaluated from the canonical CHE recurrence through the large-order connection limit of Sch\"afke and Lutz--Sch\"afke \cite{SIAM15:253:1984,CVTA34:145:1997}. The resulting spectral function $\mathcal{W}(\Omega,h)$ is analytically continued across the surface $h^2=\mathcal{M}-\Omega^2$: its zeros on the radiative sheet impose the QNM condition, while the corresponding zeros on the bound sheet impose the QBS condition. The central aim is therefore a unified treatment of the radiative and bound spectral sectors through the same global Wronskian, without enlarging the construction to boundary conditions beyond the two sectors considered here.

We test the construction in several complementary ways. The radiative-sheet roots are compared with published weak-field Ernst QNMs and with the Schwarzschild limit \cite{PLB659:375:2008,EPJC75:391:2015,EPJC82:897:2022}. On the bound sheet, the fixed-$m$ effective-mass correspondence allows direct comparison with published massive-scalar Schwarzschild QBS frequencies \cite{PRD89:083006:2014}. We also examine the local Heun-series termination prescription used in Ref.~\cite{EPJC84:57:2024} by checking the complete polynomiality criterion~\cite{JPA43:035203:2010} and the independent global Wronskian condition. All results are restricted to the separated $O(B_0^2)$ model and to the generic nonthreshold, nonresonant connection problem; no claim of spectral completeness is made.

The paper is organized as follows. Section~\ref{sec:radial_che} derives the separated weak-field radial equation and its canonical confluent-Heun form. Section~\ref{sec:connection_coefficient} constructs the endpoint-selected Wronskian and its recurrence representation. Sections~\ref{sec:qnm} and \ref{sec:qbs} apply the same spectral function to QNMs and QBSs, respectively. Section~\ref{sec:discussion} discusses the interpretation, relation to other methods, scope, and extensions of the construction, and Sec.~\ref{sec:conclusions} summarizes the main conclusions.

\section{Scalar radial equation in weakly magnetized Ernst--Schwarzschild spacetime}
\label{sec:radial_che}

We consider a massless scalar field on the magnetized Schwarzschild solution of Ernst and use units $G=c=\hbar=1$. In Schwarzschild-like coordinates, the line element is \cite{JMP17:54:1976}
\begin{equation}
 ds^2=\Lambda^2\left[-f\,dt^2+f^{-1}dr^2+r^2d\theta^2\right]
 +\frac{r^2\sin^2\theta}{\Lambda^2}d\phi^2,
 \label{eq:ernst_metric}
\end{equation}
where
\begin{equation}
 f=1-\frac{r_s}{r},
 \qquad
 \Lambda=1+\frac{1}{4}B_0^2r^2\sin^2\theta,
 \qquad
 r_s=2M.
 \label{eq:ernst_functions}
\end{equation}
Here $M$ is the black-hole mass and $B_0$ fixes the strength of the external magnetic field in the convention of Eq.~\eqref{eq:ernst_functions}.

The exact Ernst geometry is not asymptotically flat, and its scalar wave equation is not separable in spherical harmonics in general. In the weak-field treatment, terms beyond order $B_0^2$ are neglected. At this order the radial and angular sectors separate, and the radial equation acquires the effective magnetic mass term familiar from earlier analyses of scalar perturbations of magnetized Schwarzschild black holes \cite{PLB659:375:2008,EPJC75:391:2015}. Throughout this work, ``weakly magnetized'' refers specifically to this $O(B_0^2)$ separated wave equation. All spectral results below therefore refer to this separated weak-field radial model rather than to the exact nonseparable Ernst geometry.

The massless Klein--Gordon equation is
\begin{equation}
 \frac{1}{\sqrt{-g}}\partial_\mu\!\left(\sqrt{-g}\,g^{\mu\nu}\partial_\nu\Phi\right)=0.
 \label{eq:kg}
\end{equation}
With the time dependence $e^{-i\omega t}$, the weak-field equation separates with
\begin{equation}
 \Phi(t,r,\theta,\phi)=e^{-i\omega t}R(r)Y_{\ell m}(\theta,\phi).
 \label{eq:separation}
\end{equation}
At this order the angular dependence is diagonal in the spherical-harmonic basis. The angular Laplacian supplies the eigenvalue $\ell(\ell+1)$, while the azimuthal dependence contained in $Y_{\ell m}$ generates the magnetic contribution proportional to $B_0^2m^2$. The radial function therefore satisfies \cite{PLB659:375:2008,EPJC75:391:2015,EPJC84:57:2024}
\begin{equation}
 \frac{d}{dr}\left(r^2f\frac{dR}{dr}\right)
 +\left[\frac{\omega^2r^2}{f}-B_0^2m^2r^2-\ell(\ell+1)\right]R=0.
 \label{eq:radial_dimensional}
\end{equation}
Although the scalar field is massless, the combination $B_0^2m^2$ acts as an effective mass scale in the separated radial dynamics. It vanishes in the axisymmetric sector $m=0$, which reduces to the corresponding Schwarzschild scalar equation at this order.

To expose the singularity structure, we introduce the dimensionless radius $\rho=r/r_s$ together with
\begin{equation}
 \Omega=\omega r_s,
 \qquad
 \mathcal{M}=B_0^2m^2r_s^2.
 \label{eq:dimensionless_parameters}
\end{equation}
The exterior region is $\rho\geq1$, and Eq.~\eqref{eq:radial_dimensional} becomes
\begin{align}
 \frac{d^2R}{d\rho^2}
 &+\left(\frac{1}{\rho}+\frac{1}{\rho-1}\right)\frac{dR}{d\rho}
 \nonumber\\
 &+\left[
 \frac{\Omega^2}{(\rho-1)^2}
 +\frac{2\Omega^2-\mathcal{M}-\ell(\ell+1)}{\rho-1}
 \right.\nonumber\\
 &\left.\hspace{1.5em}
 +\frac{\ell(\ell+1)}{\rho}
 +\Omega^2-\mathcal{M}
 \right]R=0.
 \label{eq:radial_rho}
\end{align}
Equation~(\ref{eq:radial_rho}) is the dimensionless form of the weak-field scalar radial equation previously used in studies of magnetized Schwarzschild perturbations \cite{PLB659:375:2008,EPJC75:391:2015,EPJC84:57:2024}, up to conventions for the magnetic-field parameter. In the variables of Ref.~\cite{EPJC84:57:2024}, the same equation is obtained after the shift $\rho=1+x$. Its singularity structure determines the corresponding Heun class: $\rho=0$ and $\rho=1$ are regular singular points, whereas $\rho=\infty$ is a rank-one irregular singular point for $\Omega^2\neq\mathcal{M}$. The event horizon is located at $\rho=1$, and only $\rho=1$ and $\rho=\infty$ bound the physical exterior domain; the second regular singular point at $\rho=0$ lies in the black-hole interior.

To separate the dominant local behaviors from the remaining radial dependence, we factor out the Frobenius power $(\rho-1)^b$ at the horizon and an exponential factor $e^{h\rho}$ at infinity, writing
\begin{equation}
 R(\rho)=e^{h\rho}(\rho-1)^b y(\rho).
 \label{eq:radial_factorization}
\end{equation}
Cancellation of the constant term at infinity and of the double pole at the horizon requires
\begin{equation}
 h^2=\mathcal{M}-\Omega^2,
 \qquad
 b^2=-\Omega^2.
 \label{eq:h_b_conditions}
\end{equation}
The first relation follows from the leading constant term as $\rho\to\infty$, whereas the second follows from the coefficient of $(\rho-1)^{-2}$ in the near-horizon equation. The two horizon indices are $b=\pm i\Omega$. For the time convention in Eq.~\eqref{eq:separation}, the ingoing branch is
\begin{equation}
 b=-i\Omega.
 \label{eq:ingoing_index}
\end{equation}
Both signs of $h$ are retained at this stage, since they correspond to the two independent exponential behaviors at infinity.

Substitution of Eq.~\eqref{eq:radial_factorization} into Eq.~\eqref{eq:radial_rho} gives the confluent-Heun equation (CHE) \cite{FRANK2010} in the form
\begin{equation}
 y''+
 \left[
 \alpha+\frac{1}{\rho}+\frac{1-\mu_1}{\rho-1}
 \right]y'
 +\frac{\beta_0+\beta_1\rho}{\rho(\rho-1)}y=0.
 \label{eq:che_canonical}
\end{equation}
This is the canonical CHE specialized to the present sector $\mu_0=0$. The remaining parameters are
\begin{equation}
 \alpha=2h,
 \qquad
 \mu_1=-2b,
 \label{eq:che_parameters_1}
\end{equation}
and
\begin{equation}
 \beta_0=-\ell(\ell+1)+b-h,
 \label{eq:che_beta0}
\end{equation}
\begin{equation}
 \beta_1=2\Omega^2-\mathcal{M}+2bh+2h.
 \label{eq:che_beta1}
\end{equation}
Here $\alpha=2h$ fixes the exponential scale associated with the irregular singularity at infinity, while $\mu_1=-2b$ is fixed by the horizon Frobenius index. The coefficients $\beta_0$ and $\beta_1$ carry the remaining angular, frequency, and magnetic dependence. The equivalent Maple-type confluent-Heun parametrization used in Ref.~\cite{EPJC84:57:2024} follows after the affine transformation $z=1-\rho$, which places the event horizon at $z=0$ and the second regular singular point at $z=1$.

For $h\neq0$, the two independent radial channels at large $\rho$ follow from a direct asymptotic balance of Eq.~\eqref{eq:radial_rho}. An ansatz of the form $R\sim e^{s\rho}\rho^q$ first gives $s=\pm h$, while the next order fixes the corresponding algebraic powers of $\rho$. With unit leading coefficients, the two channels can be chosen as
\begin{equation}
 R^{(+)}_{\infty}(\rho)
 \sim e^{h\rho}\rho^{-1+(\mathcal{M}-2\Omega^2)/(2h)},
 \label{eq:asymptotic_plus}
\end{equation}
\begin{equation}
 R^{(-)}_{\infty}(\rho)
 \sim e^{-h\rho}\rho^{-1-(\mathcal{M}-2\Omega^2)/(2h)}.
 \label{eq:asymptotic_minus}
\end{equation}
At $h=0$, equivalently $\Omega^2=\mathcal{M}$, the two exponential behaviors cease to be distinct and the generic large-$\rho$ asymptotic form changes. The threshold therefore requires a separate limiting treatment. Away from this threshold, the weak-field scalar equation is reduced to a CHE with a fixed ingoing horizon branch and two well-defined independent asymptotic channels.

\section{Wronskian formulation of the global connection problem}
\label{sec:connection_coefficient}

For the ingoing horizon index fixed in Eq.~\eqref{eq:ingoing_index}, let $y_{\rm H}(\rho)$ denote the solution of Eq.~\eqref{eq:che_canonical} that is analytic at $\rho=1$ and normalized by
\begin{equation}
 y_{\rm H}(1)=1.
 \label{eq:yH_normalization}
\end{equation}
At the rank-one irregular point at infinity, the canonical asymptotic solutions depend on the Stokes sector in which the limit is taken \cite{SIAM15:253:1984,CVTA34:145:1997}. For a fixed sector and $\alpha\neq0$, we choose a unit-leading pair
\begin{equation}
 Y_{\rm alg}(\rho)
 \sim \rho^{\sigma_1}\left[1+O(\!\rho^{-1})\right],
 \label{eq:Yalg}
\end{equation}
\begin{equation}
 Y_{\rm exp}(\rho)
 \sim e^{-\alpha\rho}\rho^{\sigma_2}
 \left[1+O(\!\rho^{-1})\right].
 \label{eq:Yexp}
\end{equation}
The label ``alg'' identifies the channel whose leading behavior is purely algebraic, whereas ``exp'' denotes the independent channel carrying the additional exponential factor $e^{-\alpha\rho}$.

The powers $\sigma_1$ and $\sigma_2$ follow directly from the large-$\rho$ balance of the CHE. Substituting $y\sim\rho^\sigma$ into Eq.~\eqref{eq:che_canonical}, the leading terms of order $\rho^{\sigma-1}$ give
\begin{equation}
 \alpha\sigma_1+\beta_1=0.
 \label{eq:sigma1_balance}
\end{equation}
For the exponential channel, the ansatz $y\sim e^{-\alpha\rho}\rho^\sigma$ instead gives
\begin{equation}
 -\alpha\left(\sigma_2+2-\mu_1\right)+\beta_1=0.
 \label{eq:sigma2_balance}
\end{equation}
Hence
\begin{equation}
 \sigma_1=-\frac{\beta_1}{\alpha},
 \qquad
 \sigma_2=\mu_1-2+\frac{\beta_1}{\alpha}.
 \label{eq:sigma12}
\end{equation}
Using Eqs.~\eqref{eq:che_parameters_1} and \eqref{eq:che_beta1}, these exponents become
\begin{equation}
 \sigma_1=-b-1+\frac{\mathcal{M}-2\Omega^2}{2h},
 \label{eq:sigma1_physical}
\end{equation}
\begin{equation}
 \sigma_2=-b-1-\frac{\mathcal{M}-2\Omega^2}{2h}.
 \label{eq:sigma2_physical}
\end{equation}
After restoring the prefactor in Eq.~\eqref{eq:radial_factorization}, $Y_{\rm alg}$ and $Y_{\rm exp}$ correspond to the radial channels $R_\infty^{(+)}$ and $R_\infty^{(-)}$ in Eqs.~\eqref{eq:asymptotic_plus} and \eqref{eq:asymptotic_minus}, respectively.

Because Eqs.~\eqref{eq:Yalg} and \eqref{eq:Yexp} form a canonical basis in the chosen Stokes sector, the horizon-normalized solution has the expansion
\begin{equation}
 y_{\rm H}(\rho)=A(\Omega,h)Y_{\rm alg}(\rho)
 +B(\Omega,h)Y_{\rm exp}(\rho).
 \label{eq:connection_decomposition}
\end{equation}
The coefficients $A$ and $B$ are the regular-to-irregular connection coefficients of the selected horizon solution. With the unit-leading normalization adopted above, $A$ multiplies the radial channel $R_\infty^{(+)}$, while $B$ multiplies $R_\infty^{(-)}$. The spectral boundary conditions considered here select solutions for which the latter channel is absent. Its interpretation depends on the sheet of $h^2=\mathcal{M}-\Omega^2$, but in both spectral sectors the relevant zero condition is therefore carried by $B(\Omega,h)$. This is the reason for determining $B$ directly rather than reconstructing the complete pair of connection coefficients.

The connection decomposition can be written directly for the original radial function. Multiplying Eq.~\eqref{eq:connection_decomposition} by the prefactor in Eq.~\eqref{eq:radial_factorization} restores the radial solution associated with $y_{\rm H}$, which we denote by $R_{\rm H}$. The same prefactor maps $Y_{\rm alg}$ and $Y_{\rm exp}$ onto the unit-leading radial channels $R_\infty^{(+)}$ and $R_\infty^{(-)}$ defined by Eqs.~\eqref{eq:asymptotic_plus} and \eqref{eq:asymptotic_minus}. Therefore Eq.~\eqref{eq:connection_decomposition} becomes
\begin{equation}
 R_{\rm H}(\rho)=A(\Omega,h)R_\infty^{(+)}(\rho)
 +B(\Omega,h)R_\infty^{(-)}(\rho).
 \label{eq:radial_connection_decomposition}
\end{equation}
Thus $R_{\rm H}$ is fixed by the ingoing horizon solution, whereas $R_\infty^{(+)}$ and $R_\infty^{(-)}$ are fixed by the two asymptotic normalizations already specified in Sec.~\ref{sec:radial_che}.

The radial equation itself fixes the Wronskian of the two asymptotic solutions. With the convention $W[f,g]=fg'-f'g$, the coefficient of $R'$ in Eq.~\eqref{eq:radial_rho} is $1/\rho+1/(\rho-1)$. Abel's identity therefore gives
\begin{equation}
 \frac{d}{d\rho}\!\left\{\rho(\rho-1)
 W[R_\infty^{(+)},R_\infty^{(-)}]\right\}=0.
 \label{eq:radial_abel}
\end{equation}
Hence the quantity inside braces is independent of $\rho$. Its value is fixed by the unit-leading asymptotic forms in Eqs.~\eqref{eq:asymptotic_plus} and \eqref{eq:asymptotic_minus}, which give
\begin{equation}
 \rho(\rho-1)W[R_\infty^{(+)},R_\infty^{(-)}]=-2h.
 \label{eq:radial_basis_wronskian}
\end{equation}
Taking the Wronskian of Eq.~\eqref{eq:radial_connection_decomposition} with $R_\infty^{(+)}$ eliminates the term proportional to $A$. Using Eq.~\eqref{eq:radial_basis_wronskian}, we obtain
\begin{equation}
 \rho(\rho-1)W[R_\infty^{(+)},R_{\rm H}]
 =-2h\,B(\Omega,h).
 \label{eq:wronskian_B_relation}
\end{equation}
For $h\neq0$, the condition $B(\Omega,h)=0$ is therefore equivalent to the vanishing of this Wronskian, or equivalently to linear dependence between the horizon-selected solution and the $R_\infty^{(+)}$ channel.

It is convenient to denote the $\rho$-independent left-hand side of Eq.~\eqref{eq:wronskian_B_relation} by the global spectral Wronskian
\begin{equation}
 \mathcal{W}(\Omega,h)\equiv
 \rho(\rho-1)W[R_\infty^{(+)},R_{\rm H}].
 \label{eq:spectral_wronskian}
\end{equation}
Equation~\eqref{eq:wronskian_B_relation} then gives
\begin{equation}
 \mathcal{W}(\Omega,h)=-2h\,B(\Omega,h).
 \label{eq:spectral_wronskian_B}
\end{equation}
We now seek an algebraic representation of $\mathcal{W}$ in terms of the recurrence data of the horizon-regular CHE solution. For this purpose, introduce the Jaff\'e coordinate
\begin{equation}
 t=\frac{\rho-1}{\rho},
 \qquad
 \rho=\frac{1}{1-t},
 \label{eq:jaffe_coordinate}
\end{equation}
which maps the horizon to $t=0$ and infinity to $t=1$. The algebraic power of Eq.~\eqref{eq:Yalg} is removed by defining
\begin{equation}
 v(t)=\rho^{-\sigma_1}y_{\rm H}(\rho)
 =\sum_{k=0}^{\infty}d_k t^k.
 \label{eq:jaffe_series}
\end{equation}
Since $\rho=1$ at the expansion point, Eq.~\eqref{eq:yH_normalization} gives $d_0=1$.

Substituting Eq.~\eqref{eq:jaffe_coordinate} and the factorization $y_{\rm H}=\rho^{\sigma_1}v$ into the canonical CHE, Eq.~\eqref{eq:che_canonical}, and using $\alpha\sigma_1+\beta_1=0$, gives
\begin{align}
 &t(1-t)^2v''
 +\Big[(1-\mu_1)
 +(\alpha+\mu_1+2\sigma_1-2)t
 \nonumber\\
 &\hspace{2.4em}
 +(1-2\sigma_1)t^2\Big]v'
 \nonumber\\
 &\quad+\Big[\beta_0-\alpha\sigma_1-\mu_1\sigma_1+\sigma_1
 +\sigma_1^2 t\Big]v=0.
 \label{eq:transformed_v_equation}
\end{align}
Inserting the series in Eq.~\eqref{eq:jaffe_series} into Eq.~\eqref{eq:transformed_v_equation} yields the three-term recurrence
\begin{equation}
 (k+1)(k+1-\mu_1)d_{k+1}
 =\Phi_0(k)d_k+\Phi_1(k-1)d_{k-1},
 \label{eq:che_jaffe_recurrence}
\end{equation}
valid for $k\geq0$ with $d_{-1}=0$. The recurrence coefficients are
\begin{align}
 \Phi_0(k)={}&2k^2
 -k(\mu_1+2\sigma_1+\alpha)
 \nonumber\\
 &+\alpha\sigma_1-\beta_0+(\mu_1-1)\sigma_1,
 \label{eq:Phi0}
\end{align}
\begin{equation}
 \Phi_1(k)=-(k-\sigma_1)^2.
 \label{eq:Phi1}
\end{equation}
Thus the horizon normalization and the canonical CHE parameters determine the sequence $\{d_k\}$ successively through a minimal three-term recurrence.

The large-order behavior of this sequence carries the information needed to evaluate the global Wronskian. The Jaff\'e series is centered at the regular singular point $t=0$, whereas its large-order coefficients probe the endpoint $t=1$ associated with the irregular singularity at infinity. Using $\rho=(1-t)^{-1}$ in Eq.~\eqref{eq:connection_decomposition} together with Eqs.~\eqref{eq:Yalg} and \eqref{eq:Yexp}, the function defined in Eq.~\eqref{eq:jaffe_series} has, as $t\to1$, the asymptotic form
\begin{equation}
 v(t)\sim A(\Omega,h)+B(\Omega,h)e^{-\alpha/(1-t)}(1-t)^{-\chi}.
 \label{eq:v_near_irregular}
\end{equation}
where $\chi\equiv\sigma_2-\sigma_1=\mu_1-2+2\beta_1/\alpha=(2\Omega^2-\mathcal{M})/h$. Thus $B(\Omega,h)$ is the amplitude of the irregular exponential contribution at the compactified endpoint. For generic nonresonant parameters and a fixed Stokes sector, the regular-to-irregular theorem of Sch\"afke and Lutz--Sch\"afke relates this amplitude directly to the large-order behavior of the Jaff\'e coefficients $d_k$ \cite{SIAM15:253:1984,CVTA34:145:1997}, yielding
\begin{equation}
 \lim_{k\to\infty}
 e^{-2\sqrt{-\alpha}\sqrt{k}}k^{3/4-\chi/2}d_k
 =\frac{e^{-\alpha/2}(-\alpha)^{1/4-\chi/2}}
 {2\sqrt{\pi}}\,B(\Omega,h).
 \label{eq:schafke_large_order}
\end{equation}
Substituting Eq.~\eqref{eq:spectral_wronskian_B} into the Sch\"afke relation, Eq.~\eqref{eq:schafke_large_order}, gives the global spectral Wronskian directly in terms of the recurrence coefficients,
\begin{equation}
 \mathcal{W}(\Omega,h)=
 -4h\sqrt{\pi}\,e^{\alpha/2}
 (-\alpha)^{\chi/2-1/4}
 \lim_{k\to\infty}\mathcal{S}_k(\Omega,h).
 \label{eq:wronskian_connection}
\end{equation}
where
\begin{equation}
 \mathcal{S}_k(\Omega,h)=
 e^{-2\sqrt{-\alpha}\sqrt{k}}
 k^{3/4-\chi/2}d_k.
 \label{eq:scaled_sequence}
\end{equation}
The square root and fractional power in Eqs.~\eqref{eq:schafke_large_order}--\eqref{eq:scaled_sequence} are fixed consistently with the branch used for the canonical asymptotic basis in the chosen Stokes sector. The threshold $\alpha=2h=0$ is outside the generic construction, and resonant parameter values require the corresponding limiting prescription. Away from these exceptional cases, Eq.~\eqref{eq:wronskian_connection} evaluates the global Wronskian of the endpoint-selected radial solutions entirely from the minimal three-term recurrence.

\section{Quasinormal modes}
\label{sec:qnm}

Quasinormal modes are selected by ingoing behavior at the event horizon and outgoing behavior at infinity. The horizon condition has already been fixed by the choice $b=-i\Omega$ in Eq.~\eqref{eq:ingoing_index}. To identify the outgoing asymptotic channel, we introduce $p=\sqrt{\Omega^2-\mathcal{M}}$ on the radiative sheet defined by analytic continuation from real $\Omega>\sqrt{\mathcal{M}}$ with $p>0$, and choose $h=+ip$. On the real-frequency edge of this sheet, Eqs.~\eqref{eq:asymptotic_plus} and \eqref{eq:asymptotic_minus} contain the factors $e^{+ip\rho}$ and $e^{-ip\rho}$, respectively. With the time dependence $e^{-i\omega t}$, $R_\infty^{(+)}$ is therefore outgoing and $R_\infty^{(-)}$ incoming. This identification is preserved by analytic continuation to the complex QNM frequencies considered below.

The QNM boundary condition removes the incoming coefficient $B(\Omega,h)$ in Eq.~\eqref{eq:radial_connection_decomposition}. Since the global Wronskian satisfies Eq.~\eqref{eq:spectral_wronskian_B}, the spectral condition away from the threshold $h=0$ is
\begin{equation}
 \mathcal{W}(\Omega,+ip)=0
 \quad\Longleftrightarrow\quad
 B(\Omega,+ip)=0.
 \label{eq:qnm_condition}
\end{equation}
The vanishing of Eq.~\eqref{eq:qnm_condition} makes the horizon-selected solution linearly dependent on the outgoing channel $R_\infty^{(+)}$, so the two QNM endpoint conditions are satisfied simultaneously.

Numerically, $\mathcal{W}$ is evaluated from the recurrence representation in Eq.~\eqref{eq:wronskian_connection}. For fixed $\ell$ and $\mathcal{M}$, the radiative branch fixes $h$ and hence the CHE parameters, while the minimal three-term recurrence in Eq.~\eqref{eq:che_jaffe_recurrence} generates the coefficients $d_k$. At finite truncation $N$, the large-order limit in Eq.~\eqref{eq:wronskian_connection} is approximated by $\mathcal{S}_N$ from Eq.~\eqref{eq:scaled_sequence}. The roots reported below use $N=500$ and arbitrary-precision arithmetic and were checked to remain stable when both $N$ and the working precision are increased.

A direct comparison can be made with the continued-fraction results of Wu and Xu \cite{EPJC75:391:2015}, who use a magnetic parameter that we denote by $B_{\rm WX}$ to distinguish it from the connection coefficient $B(\Omega,h)$. Their massless-scalar effective mass is $2B_{\rm WX}|m|$, so the convention of Eq.~\eqref{eq:radial_dimensional} gives $B_0=2B_{\rm WX}$. With $M=1$ in Ref.~\cite{EPJC75:391:2015}, the dimensionless quantities are related by $\Omega=2\omega$ and $\mathcal{M}=16B_{\rm WX}^2m^2$. Table~\ref{tab:qnm_benchmark} shows representative points spanning the three mode families tabulated there. The complete set of 18 published points was also checked; the largest absolute difference is $1.23\times10^{-6}$ in $\Omega$, consistent with the six-decimal precision of the published frequencies.

\begin{table*}[t]
\centering
\caption{Representative QNM benchmark against Table~1 of Wu and Xu \cite{EPJC75:391:2015}. Their frequencies have been converted to the present convention through $\Omega=2\omega$, and $\mathcal{M}=16B_{\rm WX}^2m^2$.}
\label{tab:qnm_benchmark}
%\small
%\setlength{\tabcolsep}{5.5pt}
\begin{tabular}{ccccc}
\hline
$(\ell,m)$ & $B_{\rm WX}$ & $\mathcal{M}$ & $\Omega$ (this work) & $2\omega$ \cite{EPJC75:391:2015} \\
\hline
$(1,1)$ & 0.005 & 0.0004 & $0.585961677-0.195266311i$ & $0.585962-0.195266i$ \\
$(1,1)$ & 0.050 & 0.0400 & $0.594831322-0.189914147i$ & $0.594832-0.189914i$ \\
$(1,1)$ & 0.125 & 0.2500 & $0.642397033-0.160079241i$ & $0.642398-0.160080i$ \\
$(2,1)$ & 0.005 & 0.0004 & $0.967350872-0.193495920i$ & $0.967350-0.193496i$ \\
$(2,1)$ & 0.050 & 0.0400 & $0.973607504-0.191349163i$ & $0.973608-0.191350i$ \\
$(2,1)$ & 0.125 & 0.2500 & $1.007024342-0.179781919i$ & $1.007024-0.179782i$ \\
$(2,2)$ & 0.005 & 0.0016 & $0.967540262-0.193431019i$ & $0.967540-0.193432i$ \\
$(2,2)$ & 0.050 & 0.1600 & $0.992653212-0.184778333i$ & $0.992654-0.184778i$ \\
$(2,2)$ & 0.125 & 1.0000 & $1.129873824-0.135250399i$ & $1.129874-0.135250i$ \\
\hline
\end{tabular}
\end{table*}
%\FloatBarrier

The earlier weak-field Ernst calculation of Konoplya and Fontana \cite{PLB659:375:2008} provides a further comparison based on Frobenius, time-domain, and WKB calculations. For $M=1$, $\ell=m=1$, and their magnetic parameter $B_{\rm KF}=0.05$, corresponding to $\mathcal{M}=0.04$ in the present convention, we obtain $\omega=\Omega/2=0.297415661-0.094957074i$. Their Frobenius value is $0.297416-0.094957i$, while their independent time-domain and sixth-order WKB checks give approximately $0.295-0.096i$ and $0.2974-0.0951i$, respectively. The increase of the oscillation frequency and the simultaneous reduction of the damping magnitude as the magnetic scale grows are therefore consistent across the earlier weak-field calculations and the present Wronskian formulation.

As an additional consistency check, we finally consider the Schwarzschild limit $\mathcal{M}=0$. In this case Eq.~\eqref{eq:radial_rho} reduces to the standard massless-scalar Schwarzschild radial equation. Table~\ref{tab:qnm_schwarzschild} compares six modes obtained from Eq.~\eqref{eq:qnm_condition} with the numerical Schwarzschild spectrum reported by Mamani et al.~\cite{EPJC82:897:2022}. Their frequencies are quoted in units of $M\omega$; the reference values in the table have therefore been converted to $\Omega=2M\omega$. The agreement covers both fundamental modes and overtones.

\begin{table}[h!]
\centering
\caption{Massless-scalar Schwarzschild QNM benchmark at $\mathcal{M}=0$. Reference values from Mamani et al.~\cite{EPJC82:897:2022} are converted from $M\omega$ to $\Omega=2M\omega$.}
\label{tab:qnm_schwarzschild}
%\small
%\setlength{\tabcolsep}{2.7pt}
\begin{tabular}{ccc}
\hline
$(\ell,n)$ & $\Omega$ (this work) & $2M\omega$ \cite{EPJC82:897:2022} \\
\hline
$(0,0)$ & $0.220910-0.209791i$ & $0.220910-0.209792i$ \\
$(1,0)$ & $0.585872-0.195320i$ & $0.585872-0.195320i$ \\
$(1,1)$ & $0.528897-0.612515i$ & $0.528898-0.612514i$ \\
$(2,0)$ & $0.967288-0.193518i$ & $0.967288-0.193518i$ \\
$(2,1)$ & $0.927701-0.591208i$ & $0.927702-0.591208i$ \\
$(2,2)$ & $0.861088-1.017117i$ & $0.861088-1.017116i$ \\
\hline
\end{tabular}
\end{table}

The magnetized and zero-field benchmarks thus test the same spectral condition in two complementary regimes. Within the weak-field separated model, the QNM spectrum is obtained directly from the zeros of the global spectral Wronskian on the radiative sheet.

\section{Quasibound states}
\label{sec:qbs}

Quasibound states retain the ingoing horizon condition fixed by $b=-i\Omega$, but replace the radiative condition at infinity by exponential decay. We therefore introduce $q=\sqrt{\mathcal{M}-\Omega^2}$ on the bound sheet defined by continuation from real $|\Omega|<\sqrt{\mathcal{M}}$ with $q>0$, and choose $h=-q$. The asymptotic channels in Eqs.~\eqref{eq:asymptotic_plus} and \eqref{eq:asymptotic_minus} then contain the factors $e^{-q\rho}$ and $e^{+q\rho}$, respectively. Thus $R_\infty^{(+)}$ is the decaying channel and $R_\infty^{(-)}$ the growing one. Since the latter is multiplied by $B(\Omega,h)$ in Eq.~\eqref{eq:radial_connection_decomposition}, the QBS condition away from the threshold $h=0$ is
\begin{equation}
 \mathcal{W}(\Omega,-q)=0
 \quad\Longleftrightarrow\quad
 B(\Omega,-q)=0.
 \label{eq:qbs_condition}
\end{equation}
Equations~\eqref{eq:qnm_condition} and \eqref{eq:qbs_condition} therefore select zeros of the same global spectral function on different sheets of $h^2=\mathcal{M}-\Omega^2$. 

Numerically, the QBS condition in Eq.~\eqref{eq:qbs_condition} is solved using the same recurrence evaluation of $\mathcal{W}$ as in Sec.~\ref{sec:qnm}, now on the bound sheet with $h=-q$. The results below use $N=700$ and arbitrary-precision arithmetic, and representative roots remain stable when the truncation order and working precision are increased.

An earlier confluent-Heun treatment of the same weak-field radial equation obtained a frequency condition by imposing $\delta/\alpha+(\beta+\gamma)/2=-n$, with $n=1,2,\ldots$, and identifying this relation as the polynomial condition for the local confluent-Heun function HeunC~\cite{EPJC84:57:2024}. In the standard HeunC normalization, however, a polynomial of degree $N$ is obtained if and only if two termination conditions are satisfied simultaneously \cite{JPA43:035203:2010}:
\begin{equation}
 \frac{\delta}{\alpha}+\frac{\beta+\gamma}{2}+N+1=0,
 \qquad
 \Delta_{N+1}=0.
 \label{eq:heunc_polynomial_conditions}
\end{equation}
The first relation is the $\delta_N$ condition, while the second is the finite determinant condition. Equivalently, if $\mathrm{HeunC}(z)=\sum_{j=0}^{\infty}c_j z^j$ is normalized by $c_0=1$, the second condition requires $c_{N+1}=0$ \cite{JPA43:035203:2010}. The condition used in Ref.~\cite{EPJC84:57:2024} is therefore only the first of Eq.~\eqref{eq:heunc_polynomial_conditions}, with $N=n-1$, and by itself does not establish polynomiality. The present construction does not impose polynomial truncation: the physical QBS condition is instead the global removal of the growing asymptotic channel through Eq.~\eqref{eq:qbs_condition}.

To test the consequence directly, we evaluate the frequencies $\Omega_{\rm S}$ obtained from the condition of Ref.~\cite{EPJC84:57:2024} at $\mathcal{M}=0.64$, choosing the roots with $\operatorname{Re}\Omega_{\rm S}>0$ and $\operatorname{Re}q_{\rm S}>0$. Table~\ref{tab:senjaya_qbs_test} shows that the second termination coefficient $c_{N+1}$ is nonzero for each tested pair $(\ell,n)$. Direct evaluation of Eq.~\eqref{eq:wronskian_connection} at the same frequencies also remains nonzero as the truncation is increased through $N=400,800,1200$; the global condition is therefore not satisfied in any of the four cases. Hence the tested frequencies satisfy neither the complete polynomial criterion nor the global QBS condition in Eq.~\eqref{eq:qbs_condition}.

\begin{table}[t]
\centering
\caption{Diagnostic test of the frequencies $\Omega_{\rm S}$ obtained from the single termination relation used in Ref.~\cite{EPJC84:57:2024}, at $\mathcal{M}=0.64$. The associated polynomial degree is $N=n-1$, and $c_{N+1}$ tests the second condition in Eq.~\eqref{eq:heunc_polynomial_conditions}.}
\label{tab:senjaya_qbs_test}
%\small
%\setlength{\tabcolsep}{3.0pt}
\begin{tabular}{ccc}
\hline
$(\ell,n)$ & $\Omega_{\rm S}$ & $|c_{N+1}|$ \\
\hline
$(1,1)$ & $0.78235479+0.03075941i$ & $0.580813$ \\
$(2,1)$ & $0.78235479+0.03075941i$ & $2.669292$ \\
$(1,2)$ & $0.79010997+0.00795812i$ & $0.295010$ \\
$(2,2)$ & $0.79010997+0.00795812i$ & $1.822835$ \\
\hline
\end{tabular}
\end{table}

A second check follows from the effective-mass structure of the weak-field radial equation. For a fixed azimuthal sector $m$, Eq.~\eqref{eq:radial_dimensional} is formally identical to the radial equation of a massive scalar field in Schwarzschild spacetime under the identification $\mu_{\rm eff}=B_0|m|$, or $M\mu_{\rm eff}=\sqrt{\mathcal{M}}/2$. Published massive-scalar Schwarzschild quasibound frequencies therefore provide a direct benchmark for the bound-sheet roots. Barranco et al.~\cite{PRD89:083006:2014} tabulate the first nine $\ell=1$ quasibound frequencies for $M\mu=0.20$ and $0.30$ in terms of Laplace frequencies $s_n$, for which $\operatorname{Re}s_n<0$ gives the decay rate and $\operatorname{Im}s_n>0$ the oscillation frequency. In the present $e^{-i\omega t}$ convention, their values are converted through $M\omega=\operatorname{Im}(Ms_n)+i\operatorname{Re}(Ms_n)$. Table~\ref{tab:qbs_massive_schwarzschild} compares the first three modes at each coupling with the zeros of Eq.~\eqref{eq:qbs_condition}. The largest absolute difference over the six entries is $9.1\times10^{-8}$ in $M\omega$, consistent with the precision quoted in the published table.

\begin{table}[t]
\centering
\caption{Comparison of the $\ell=1$ quasibound-state frequencies from Eq.~\eqref{eq:qbs_condition} with Table~1 of Barranco et al.~\cite{PRD89:083006:2014}. Their Laplace frequencies are converted through $M\omega=\operatorname{Im}(Ms_n)+i\operatorname{Re}(Ms_n)$, and $M\mu_{\rm eff}=\sqrt{\mathcal{M}}/2$.}
\label{tab:qbs_massive_schwarzschild}
%\small
%\setlength{\tabcolsep}{2.8pt}
\begin{tabular}{ccccc}
\hline
$M\mu_{\rm eff}$ & $n$ & component & this work & Ref.~\cite{PRD89:083006:2014} \\
\hline
$0.20$ & $1$ & $\operatorname{Re}(M\omega)$ & $0.1989526613$ & $0.1989527$ \\
       &     & $-\operatorname{Im}(M\omega)$ & $4.0604383\!\times\!10^{-8}$ & $4.060\!\times\!10^{-8}$ \\
$0.20$ & $2$ & $\operatorname{Re}(M\omega)$ & $0.1995374633$ & $0.19953747$ \\
       &     & $-\operatorname{Im}(M\omega)$ & $1.4735507\!\times\!10^{-8}$ & $1.473\!\times\!10^{-8}$ \\
$0.20$ & $3$ & $\operatorname{Re}(M\omega)$ & $0.1997415499$ & $0.19974155$ \\
       &     & $-\operatorname{Im}(M\omega)$ & $6.5827274\!\times\!10^{-9}$ & $6.582\!\times\!10^{-9}$ \\
$0.30$ & $1$ & $\operatorname{Re}(M\omega)$ & $0.2961923464$ & $0.2961924$ \\
       &     & $-\operatorname{Im}(M\omega)$ & $9.4556517\!\times\!10^{-6}$ & $9.4556\!\times\!10^{-6}$ \\
$0.30$ & $2$ & $\operatorname{Re}(M\omega)$ & $0.2983351096$ & $0.2983352$ \\
       &     & $-\operatorname{Im}(M\omega)$ & $3.6585517\!\times\!10^{-6}$ & $3.6585\!\times\!10^{-6}$ \\
$0.30$ & $3$ & $\operatorname{Re}(M\omega)$ & $0.2990799579$ & $0.29907996$ \\
       &     & $-\operatorname{Im}(M\omega)$ & $1.6491034\!\times\!10^{-6}$ & $1.6491\!\times\!10^{-6}$ \\
\hline
\end{tabular}
\end{table}

The two tests address complementary aspects of the bound-state problem. The first shows that, for the cases tested, the frequencies selected by the HeunC $\delta_N$ relation alone do not satisfy the global QBS condition, while the second confirms the bound-sheet Wronskian roots against independent massive-scalar Schwarzschild benchmarks.

\section{Discussion}
\label{sec:discussion}

The common Wronskian description of QNMs and QBSs can be understood directly from the boundary-value structure of the separated weak-field problem. Once the ingoing horizon solution is fixed, the remaining spectral requirement is imposed at infinity, and $\mathcal{W}(\Omega,h)=-2hB(\Omega,h)$ vanishes when the horizon-selected solution becomes linearly dependent on the asymptotic solution required by that boundary condition. On the radiative sheet the selected asymptotic behavior is outgoing, whereas analytic continuation to the bound sheet selects the decaying solution. The change from a QNM to a QBS condition is therefore carried by the branch of $h^2=\mathcal{M}-\Omega^2$ and the associated asymptotic interpretation, rather than by introducing a second quantization mechanism. In this restricted sense, the same endpoint-defined spectral object provides a unified treatment of the radiative and bound sectors. Its global character refers specifically to the relation between the horizon and infinity for these two endpoint selections, not to a construction of all possible connection data.

The present construction belongs to a broader development in which confluent-Heun connection data are used directly in black-hole spectral and scattering problems \cite{PRD105:044047:2022,PRD112:103036:2025}. The closest methodological comparison is with Chen et al.~\cite{PRD112:103036:2025}. Relative to that work, the physical scope considered here is narrower, being restricted to the QNM and QBS sectors of the weak-field Ernst scalar equation. The methodological point of contact is limited to the use of confluent-Heun connection information to impose boundary conditions across distinct asymptotic regions; the actual constructions and spectral objects are different. Here the spectral condition is defined directly as the Wronskian of the horizon-selected and infinity-selected solutions and is evaluated from the problem-specific CHE recurrence through the large-order connection limit of Sch\"afke and Lutz--Sch\"afke \cite{SIAM15:253:1984,CVTA34:145:1997}. Accordingly, the comparison with Chen et al. concerns the organization of global connection information rather than an equivalence between the two methods or a competition between numerical algorithms. 

The magnetic parameter acquires a direct physical interpretation within this separated weak-field description. For each fixed azimuthal number $m$, the radial equation is formally identical to that of a massive scalar in Schwarzschild spacetime under the identification $\mu_{\rm eff}=B_0|m|$ \cite{PLB659:375:2008,EPJC75:391:2015}. This mode-by-mode equivalence explains why increasing the magnetic scale produces spectral behavior familiar from the massive-scalar problem and why published massive-Schwarzschild QBS frequencies provide an independent benchmark for the bound-sheet calculation. The correspondence is nevertheless limited to the radial equation at fixed $m$. The effective mass changes from one azimuthal sector to another and vanishes for $m=0$, so the magnetized scalar problem cannot be identified globally with a single massive field. The magnetic origin of the scale remains essential when the full set of azimuthal sectors is considered.

The endpoint formulation also separates the physical QBS condition from confluent-Heun polynomiality. A QBS is selected by compatibility between the ingoing horizon solution and the decaying asymptotic channel, encoded globally by $\mathcal{W}=0$, whereas polynomiality asks whether the local CHE series terminates. The diagnostic of Sec.~\ref{sec:qbs} makes this distinction explicit: in the representative cases tested, satisfying the single local termination relation used in Ref.~\cite{EPJC84:57:2024} is not sufficient to ensure the complete polynomiality criterion \cite{JPA43:035203:2010} or the endpoint-selected QBS boundary condition. Polynomial solutions may coincide with physical QBSs only when the full termination conditions are simultaneously compatible with the boundary conditions at both endpoints; local truncation alone does not impose the global bound-state problem.

The interpretation of these results must remain tied to the approximations under which the separated radial problem is obtained. The exact Ernst--Schwarzschild spacetime is not asymptotically flat \cite{JMP17:54:1976,PRD113:124029:2026}. For scalar perturbations of the exact magnetized geometry, a spherical-harmonic decomposition leads in general to a coupled multipolar system rather than to a single separated radial equation, as shown explicitly for the charged-scalar problem by Ribeiro et al.~\cite{PRD113:124029:2026}. The spectra discussed here therefore belong specifically to the $O(B_0^2)$ separated model employed in Refs.~\cite{PLB659:375:2008,EPJC75:391:2015} and should not be identified with the spectrum of the unrestricted exact-Ernst problem. The condition $h\neq0$ has a similarly concrete meaning for the connection construction. At the threshold $\Omega^2=\mathcal{M}$ the two exponential asymptotic behaviors coalesce, and the generic decomposition used to define the regular-to-irregular connection coefficient must be replaced by an appropriate limiting analysis. Resonant CHE parameters require the analogous care because the generic connection formula is not directly applicable there. In addition, the numerical tests establish the behavior of the branches and overtones that were explicitly followed, but they do not prove that all spectral families have been exhausted. These limitations indicate what must change in any extension of the present analysis: threshold and resonant cases require dedicated limiting connection formulas, while going beyond the $O(B_0^2)$ regime requires a formulation capable of treating the coupled and non-asymptotically-flat structure of the full Ernst geometry. Within problems that retain a controlled confluent-Heun reduction, the same endpoint-selected Wronskian logic can be tested without enlarging the present claim beyond the QNM and QBS sectors considered here.

\section{Conclusions}
\label{sec:conclusions}

We considered a massless scalar field governed by the Klein--Gordon equation in magnetized Ernst--Schwarzschild spacetime. Working in the weak-magnetization regime and retaining terms through $O(B_0^2)$, the wave equation separates in spherical harmonics and the radial problem can be reduced to a confluent-Heun equation after the dominant horizon and infinity behaviors are factored out \cite{JMP17:54:1976,PLB659:375:2008,EPJC75:391:2015}. The main aim of this work was to formulate the physical horizon-to-infinity spectral condition directly within that confluent-Heun structure. Starting from the solution fixed to be ingoing at the event horizon, we defined an endpoint-selected spectral Wronskian and evaluated it from the canonical CHE recurrence through the regular-to-irregular large-order connection limit of Sch\"afke and Lutz--Sch\"afke \cite{SIAM15:253:1984,CVTA34:145:1997}. The resulting construction turns the local Heun representation into a direct computational condition for the global radial boundary-value problem.

Within this construction, quasinormal modes and quasibound states do not require independent quantization procedures. Both are selected by zeros of the same analytic spectral function $\mathcal{W}(\Omega,h)$, evaluated on the radiative and bound sheets of $h^2=\mathcal{M}-\Omega^2$, respectively. The physical boundary conditions remain different, since one selects the outgoing asymptotic channel and the other the decaying one, but the underlying horizon-to-infinity connection condition is the same. This is the main structural result of the analysis and provides a unified treatment of the radiative and bound spectral sectors within the separated weak-field Ernst problem.

The numerical applications test this same spectral object in two complementary regimes. On the radiative sheet, the Wronskian condition reproduces the weak-field Ernst QNM results used as benchmarks and the Schwarzschild limit \cite{PLB659:375:2008,EPJC75:391:2015,EPJC82:897:2022}. On the bound sheet, it reproduces published massive-scalar Schwarzschild quasibound frequencies under the fixed-$m$ identification $\mu_{\rm eff}=B_0|m|$ \cite{PRD89:083006:2014,PLB659:375:2008,EPJC75:391:2015}. These independent checks support the analytic continuation of the recurrence-based Wronskian between the two spectral sectors. The additional polynomiality diagnostic of Sec.~\ref{sec:qbs} also shows that, for the representative cases tested, the frequencies selected by the local Heun-series termination relation employed in Ref.~\cite{EPJC84:57:2024} do not satisfy either the complete polynomiality criterion or the endpoint-selected QBS condition.

The conclusions above apply to the separated $O(B_0^2)$ model and to the generic nonthreshold, nonresonant connection problem, and they do not constitute a proof of spectral completeness. Threshold points and resonant CHE parameters require dedicated limiting constructions, while extending the analysis beyond the weak-field separated regime requires a formulation adapted to the coupled and non-asymptotically-flat structure of the full magnetized geometry \cite{PRD113:124029:2026}. Within these limits, the present results show that the analytic structure of the confluent-Heun equation can be used for more than a local representation of the radial solution: its recurrence and regular-to-irregular connection data provide a practical route to the global spectral conditions selecting both quasinormal and quasibound states.

\section*{Statements and Declarations}

\subsection*{Competing interests}
The authors declare no competing interests.

\subsection*{Data availability}
Data sharing is not applicable to this article as no datasets were generated or analyzed during the current study.

\begin{acknowledgements}
This work was supported in part by means of funds provided by CNPq, Brazil, Grant Nos. 308172\slash 2023-0 and CNPq\slash Universal\slash 420896/2025-2, FAPEMA and CAPES (Finance code 001).
\end{acknowledgements}

\bibliographystyle{spphys}       % APS-like style for physics
%\bibliography{mybibfile2020.bib}

\end{document}